\documentclass[twocolumn,aps,amssymb,floatfix]{revtex4}

\usepackage{placeins}

\usepackage{graphicx}
\usepackage{bm}
\usepackage{amsmath}
\usepackage{amssymb}
\usepackage{amsfonts}
\usepackage{float}
\usepackage{hyperref}
\usepackage{dsfont}  
\usepackage{slashed}  
\usepackage{booktabs}
\usepackage{multirow}
\usepackage{subfigure}
\usepackage[sort&compress]{natbib}

\newcommand{\cl}{\centerline}
\newcommand{\be}{\begin{equation}}
\newcommand{\ee}{\end{equation}}
\newcommand{\bea}{\begin{eqnarray}}
\newcommand{\eea}{\end{eqnarray}}

\usepackage{tikz}
\usetikzlibrary{shadows}
\usetikzlibrary{arrows,shapes,positioning}
\usetikzlibrary{decorations.markings}
\usetikzlibrary{decorations.pathmorphing}
\usetikzlibrary{decorations.pathreplacing}
\tikzstyle arrowstyle=[scale=1]
\tikzstyle directed=[postaction={decorate,decoration={markings, mark=at position .65 with {\arrow[arrowstyle]{stealth}}}}]
\tikzstyle end directed=[postaction={decorate,decoration={markings, mark=at position 1 with {\arrow[arrowstyle]{stealth}}}}]
\tikzstyle reverse directed=[postaction={decorate,decoration={markings, mark=at position .65 with {\arrowreversed[arrowstyle]{stealth};}}}]
\tikzstyle{ann} = [fill=white,font=\footnotesize,inner sep=1pt]

\begin{document}

\title{Lattice Generalized Parton Distributions and Form Factors of the Nucleon\\ [-2ex] $\quad$}

\author{Martha Constantinou}
\email{marthac@temple.edu}
\affiliation{$\quad$\\Temple University \\ 1925 N. 12th Street, Philadelphia, PA 19122-1801, USA\\$\quad$\\$\quad$}

\begin{abstract}
In these proceedings we discuss recent progress in nucleon structure 
using lattice QCD simulations at or near the physical value of the pion
mass. Main focus will be given in observables such as the nucleon
axial charge and the first moments of parton distributions, for both the
valence and sea quark contributions, and discuss their implications on
the spin content of the nucleon. We will will also report developments on 
the evaluation of the gluon momentum fraction, which contributes significantly
to the nucleon spin.

\end{abstract}

\maketitle
\bibliographystyle{apsrev}

\section{Introduction}
\label{sec1}

The distribution of the proton spin among its constituents has been a long-standing puzzle since the first
exploration of the internal structure of the proton. First measurements of the spin dependent structure
of the proton by the European Muon Collaboration~\cite{Ashman:1987hv} revealed that quarks only 
contribute about half of the proton's total spin. This was called the proton spin crisis and since then
both experimentalists and theorists have invested significant resources to understand it.  

Due to tremendous theoretical and algorithmic advances, lattice QCD can provide 
{\it{ab initio}} simulations of the quark and gluon contributions to the nucleon spin, and to provide a 
rigorous test to the theory of QCD.  The lattice QCD calculations include matrix elements of local 
operators, which can be related to form factors and generalized form factor relevant for the nucleon 
spin and thus have the potential to reveal the spin structure.

 Historically, the first studies could provide estimates for the valence quark contribution to the spin of the nucleon, with the
 sea quark contributions being intensively explored quite recently. Currently, while the computations of the quark distribution functions 
 are approaching a satisfactory situation, the case of the gluon contributions is much less advanced: first results
 using dynamical simulations only appeared within the last two years.

The algorithmic improvements and increase of available computational resources have allows simulations directly at the 
physical value of the pion mass to become feasible; this has eliminated a large uncertainty related to the uncontrolled
chiral extrapolations.

\bigskip

The total nucleon spin is generated by the quark orbital angular momentum ($L^q$), the quark spin ($\Delta\Sigma^q$) 
and the gluon angular momentum ($J^G$) via a gauge invariant decomposition~\cite{Ji:1996ek},  which may be implemented 
in the lattice formalism. In particular, the quark components are related to the axial charge ($g_A^q$) and the generalized 
form factors of the one-derivative vector at zero momentum transfer ($Q^2{=}0$):
\bea
\label{eq1}
\frac{1}{2} = \sum_q \left(L^q + \frac{1}{2}\Delta\Sigma^q \right) + J^G\,, \qquad \qquad  \\
J^q = \frac{1}{2} \left(A^q_{20} + B^q_{20} \right)\,, \quad
L^q = J^q - \Delta\Sigma^q\,,
\Delta\Sigma^q = g_A^q\,.
\label{eq2}
\eea
All quantities shown above depend on the momentum transferred squared, $Q^2$, but we are interested in the 
$Q^2{=}0$ limit.
The generalized form factors of the first unpolarized moment, $A^q_{20}$ and $B^q_{20}$, enter the
definition of the spin. $A^q_{20}(Q^2{=}0)$ is the quark momentum fraction, and unlike the case of $B^q_{20}$ 
is extracted directly from lattice data. For the estimation of $B^q_{20}(Q^2{=}0)$ one relies on fits of its momentum
dependence.

Since it is necessary to include the individual quark contributions to the various components of the spin,
we must also consider the disconnected contributions for $g_A$, $A_{20}$ and $B_{20}$. Furthermore, the 
nucleon matrix elements of the strange and charm quark operators should also be considered, which are entirely 
disconnected and relatively cheap to compute due to the larger quark mass. These will determine the strange and 
charm quark spin contributions to the nucleon spin without neglecting any parts, which will eventually provide a 
stringent check on the distribution of the spin to the various quark degrees of freedom.

\vspace*{-.25cm}
\section{Nucleons on the Lattice}
\vspace*{-.1cm}
\label{sec2}

\begin{figure}[h!]
  \centering
  \begin{tikzpicture}[scale = 0.3]

\node[ann] at (10,2) {$N(x)$};
\node[ann] at (0,2) {$\overline{N}(x')$};
\node[ann] at (5,3.2) {$\mathcal J(x_1)$};

\draw[black, directed, line width=1pt] (0,0) -- (10,0);
\draw[black, directed,line width=1pt]  (5.0,2.3) to[out=0,in=130] (10,0);
\draw[black,directed, line width=1pt] (0,0) to[out=50,in=180]   (5,2.3);
\draw[black, directed,line width=1pt] (0,0) to[out=-50,in=-130] (10,0);
\draw[black, line width=1pt] (4.7,2.6) -- (5.3,2.0);
\draw[black, line width=1pt] (5.3,2.6) -- (4.7,2.0);
\end{tikzpicture} \quad
  \begin{tikzpicture}[scale = 0.3]

\node[ann] at (10,2) {$N(x)$};
\node[ann] at (0,2) {$\overline{N}(x')$};
\node[ann] at (7,3) {$\mathcal J(x_1)$};

\draw[black, directed, line width=1pt] (0,0) -- (10,0);
\draw[black,directed, line width=1pt] (0,0) to[out=50,in=130]   (10,0);
\draw[black, directed,line width=1pt] (0,0) to[out=-50,in=-130] (10,0);

\draw[black, line width=1pt] (4.7,3.6) -- (5.3,3.0);
\draw[black, line width=1pt] (5.3,3.6) -- (4.7,3.0);

\draw[black, line width=1pt] (5.0,4.3) circle(1cm);

\end{tikzpicture} \\[-3ex] 
  \begin{tikzpicture}[scale = 0.3]

\node[ann] at (0,2) {$\overline{N}(x')$};
\node[ann] at (10,2) {$N(x)$};
\node[ann] at (7,2.8) {$\mathcal J(x_1)$};

\draw[black, directed, line width=1pt] (0,0) -- (10,0);
\draw[black,directed, line width=1pt] (0,0) to[out=50,in=130]   (10,0);
\draw[black, directed,line width=1pt] (0,0) to[out=-50,in=-130] (10,0);

\draw[black, line width=1pt] (4.7,3.6) -- (5.3,3.0);
\draw[black, line width=1pt] (5.3,3.6) -- (4.7,3.0);

\draw[decorate, rotate
around={-90:(5.0,4.3)},decoration={coil,amplitude=3pt,
segment length=3pt}] (5.0,4.3) circle(1cm);

\end{tikzpicture}
\vspace*{-0.35cm}
  \caption{\label{fig1} 
{\footnotesize{Quark and gluon contributions to the nucleon three-point function. The current
insertion ($J(x_1)$) is indicated by an {\bf{$\times$}} symbol. 
{\bf Upper Left:} connected, {\bf Upper right:} disconnected quark loop, {\bf lower:} disconnected gluon loop.}}}
\end{figure}
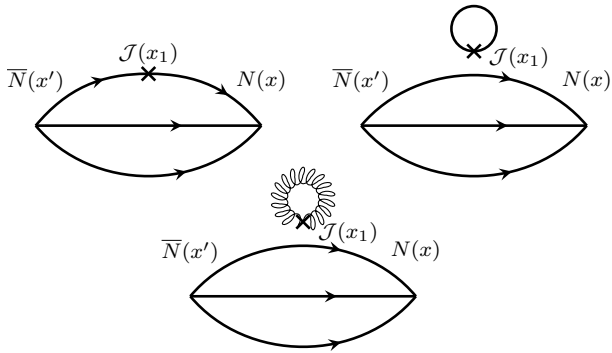

To study the quark contributions to nucleon quantities one needs to compute three-point functions, schematically represented by 
the two upper diagrams shown in Fig.~\ref{fig1}, the connected (left) and disconnected (right). The lower diagram is also disconnected and includes 
a gluonic closed loop, and corresponds to contributions from the gluonic degrees of freedom. Both disconnected diagrams have 
only been limitedly studied in the past because the signal-to-noise ratio is very small and they require special techniques in order to 
obtain results with controlled statistical and systematic uncertainties.  
Over the last few years calculations of the disconnected quark contributions have appeared in the literature
\cite{Babich:2010at,Engelhardt:2012gd,Stathopoulos:2013aci,Abdel-Rehim:2013wlz,Chambers:2015bka,Gambhir:2016jul} and quite recently 
results at the physical point have become available~\cite{Yang:2015uis,Abdel-Rehim:2016won,Abdel-Rehim:2016pjw} (see also Ref.~\cite{SaraLat16} 
for a recent review). The advances in the computation of disconnected diagrams with closed quark loops  have initiated calculations of
the diagram with the gluon loop~\cite{Yang:2016plb,Alexandrou:2016ekb}. 
\vskip -0.25cm
\FloatBarrier
For the extraction of the quantities of interest one must form dimensionless ratios of the two- and three- point correlation functions, denoted as
$G^{2pt}$ and $G^{3pt}$, respectively. The ratio is optimized so that it does not contain potentially noisy two-point functions at large time separations 
and because correlations between its different factors reduce the statistical noise.
\vskip -0.4cm
\bea
R_{\cal O}(\Gamma,\vec q, t, t_f) \hspace*{-0.2cm}&=&\hspace*{-0.2cm} \frac{G^{3pt}_{\cal O}(\Gamma,\vec q,t)}{G^{2pt}(\vec 0, t_f)}\,\times \nonumber \\
\hspace*{-0.1cm}&&\hspace*{-0.2cm}\sqrt{\frac{G^{2pt}({-}\vec q, t_f{-}t)G^{2pt}(\vec 0, t)G^{2pt}(\vec 0, t_f)}{G^{2pt}(\vec 0  , t_f{-}t)G^{2pt}({-}\vec q,t)G^{2pt}({-}\vec q,t_f)}}\,\, \nonumber\\[1ex]
&&\hspace*{-1cm}{\rightarrow \atop {{t_f{-}t\rightarrow \infty} \atop {t{-}t_i\rightarrow \infty}}}\,\,
\Pi (\Gamma,\vec q) \,.
\label{EqRatio}
\eea
\vskip -0.1cm
Details on the definition of the above quantities can be found in Ref.~\cite{Constantinou:2014tga}. 
In the aforementioned ratio, it is important to keep the time separation of the initial (source) and final (sink) states of the nucleon 
large enough to ensure suppressed contamination from excited states. The desired information may be extracted from a plateau with 
respect to the current insertion time, $t$, which is well separated from the source and the sink in order
to avoid overlap with the excited states.

Lattice data extracted from a non-conserved current must be renormalized prior a comparison with experimental data and phenomenological analyses.
For cases like the nucleon charges and the quark momentum fraction, the renormalization is multiplicative, while for the case of the
gluon momentum fraction, a more complicated renormalization prescription is required due to mixing with other operators.
Finally, the properly renormalizes matrix elements may be expressed in terms of generalized form factors, which provide
information on the nucleon structure.

\vspace*{-.2cm}
\section{Quark Contributions}
\label{sec3}
\vspace*{-.2cm}
\subsection{Axial charge}
\label{sec3sub1}
\vspace*{-.2cm}
One of the fundamental nucleon observables is the axial charge, $g_A$, which governs the rate of $\beta$-decay 
and has been measured precisely. It is essential for lattice QCD to reproduce its experimental value, or in the case of
deviation to understand its origin, so that we have confidence in predicting quantities that are not easily accessible
in experiments. The axial charge can be determined directly from lattice data without the need of fitting 
a momentum dependence, and thus, it is a benchmark quantity for hadron structure computations.
\vskip -0.15cm
\begin{figure}[!h]
\cl{\hspace*{-0.5cm}\includegraphics[scale=0.14]{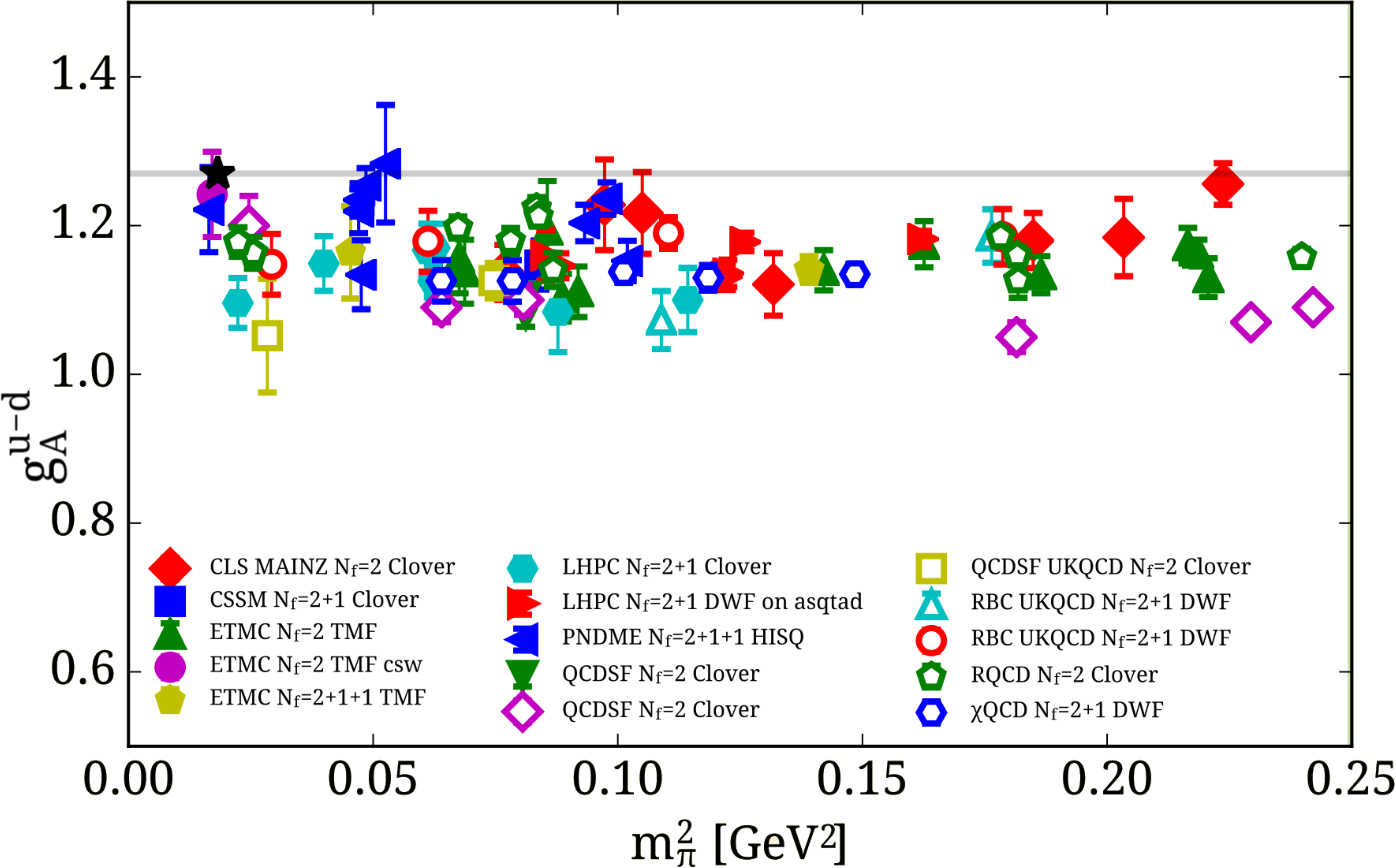}}
\vskip -0.25cm
\caption{\footnotesize{Lattice results on $g_A$ vs $m_\pi^2$ for: $N_f{=}2{+}1$ DWF
(RBC/UKQCD~\cite{Yamazaki:2008py, Yamazaki:2009zq},
RBC/UKQCD~\cite{Ohta:2013qda}, $\chi$QCD~\cite{Liu:2015nva}),
$N_f{=}2{+}1$ DWF on asqtad sea (LHPC~\cite{Bratt:2010jn}), $N_f{=}2$
TMF (ETMC~\cite{Alexandrou:2010hf}), $N_f{=}2$ Clover
(QCDSF/UKQCD~\cite{Pleiter:2011gw}, CLS/MAINZ~\cite{Capitani:2012gj},
QCDSF~\cite{Horsley:2013ayv}, RQCD~\cite{Bali:2014nma}),
$N_f{=}1{+}2$ Clover (LHPC~\cite{Green:2012rr},
CSSM~\cite{Owen:2012ts}), $N_f{=}2{+}1{+}1$ TMF
(ETMC~\cite{Alexandrou:2013joa}), $N_f{=}2{+}1{+}1$ HISQ
(PNDME~\cite{Bhattacharya:2016zcn}), $N_f{=}2$ TMF with Clover
(ETMC~\cite{Abdel-Rehim:2015owa}). The black star shows the experimental value.}}
\label{fig2}
\end{figure}
\FloatBarrier

In Fig.~\ref{fig2} we plot $g_A$ as a function of the pion mass, $m_\pi$,
for simulations with $m_\pi {\le} 500$MeV. The plotted results correspond to different
lattice spacings, volume, number of dynamical quarks and formulations: 
Clover, Domain Wall (DWF), HISQ, Staggered and Twisted Mass (TMF) fermions~\cite{Yamazaki:2008py,
  Yamazaki:2009zq, Bratt:2010jn, Alexandrou:2010hf, Pleiter:2011gw,
  QCDSF:2011aa, Capitani:2012gj, Green:2012rr, Owen:2012ts,
  Horsley:2013ayv, Alexandrou:2013joa, Bhattacharya:2013ehc,
  Ohta:2013qda, Bali:2014nma,Abdel-Rehim:2015owa}. We compare only results obtained from the plateau 
method without continuum extrapolation and volume corrections. 

Over the last years,
simulations at or near the physical point have become available, which eliminate
the uncontrolled systematic on the chiral extrapolation. We find that to the
current statistics, volume and lattice spacing, the data close to the physical pion mass
have an upward tendancy towards the experimental value: $g_A^{\rm exp}=1.2701(25)$~\cite{Beringer:1900zz}. 
However, statistical and systematic uncertainties are not well under controlled yet and it is crucial 
to increase the statistics and study the volume dependence before reaching to final conclusions.

One of the most important systematic uncertainties is the contamination by 
excited states, as the interpolating field used for the ground state also couples to 
the excited ones. The identification of the ground state for the three-point functions is 
more saddle compare to the two-point function, and different analysis techniques are used 
in order to extract reliable results. The most common approach is the plateau method 
in which one probes the large Euclidean time evolution of the ratio in Eq.~(\ref{EqRatio}), 
and the excited states contributions fall exponentially with the sink-insertion ($t_f{-}t$) and
insertion-source ($t{-}t_i$) time separation. Thus, by increasing the source-sink
separation a decrease of the excited states contamination is achieved. However,
statistical noise is increased, especially for simulations at the physical point. Alternatively, 
the matrix element may be obtained by performing 2- or 3-state fits to account for 
contributions from the first and second excited states.  A third method is the summation method 
in which one sums the ratio from the source to the sink and the excited state contaminations are 
suppressed by exponentials decaying with $(t_f{-}t_i)$ and $g_A$ is extracted from the slope
slope of the summed ratio.
\vskip -0.25cm
\begin{figure}[!h]
\cl{\includegraphics[scale=0.44]{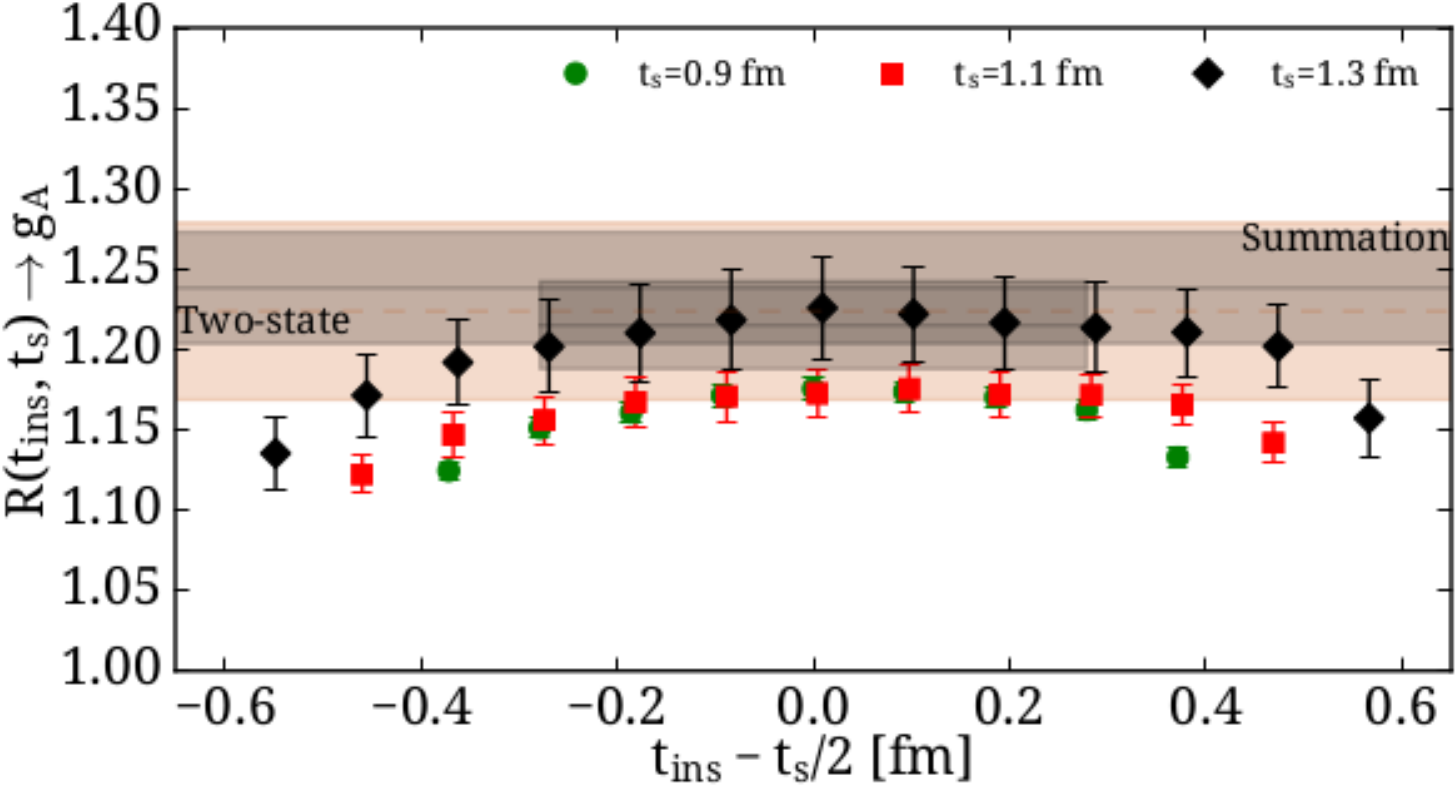}}
\vskip -0.25cm
\caption{\footnotesize{The ratio of the axial charge for $N_f{=}2$ TMF+clover~\cite{Abdel-Rehim:2015owa})
at $m_\pi{\sim}130$ MeV for tree source-sink separations. The value extracted from the 2-state fit and summation are also
shown with a pink and grey bands, respectively.}}
\label{fig3}
\end{figure}
\FloatBarrier
All three methods have been applied for the case of $g_A$ and we present recent results by
ETMC~\cite{Abdel-Rehim:2015owa} at $m_\pi{\sim} 130$ MeV, where very small change of $g_A$
has been seen by varying the source-sink separation from 0.9-1.3 fm (see Fig.~\ref{fig3}). In fact,
at a separation of 1.3 fm, the plateau, 2-state fit and summation methods are in agreement.

\begin{figure}[!h]
\cl{\includegraphics[scale=0.48]{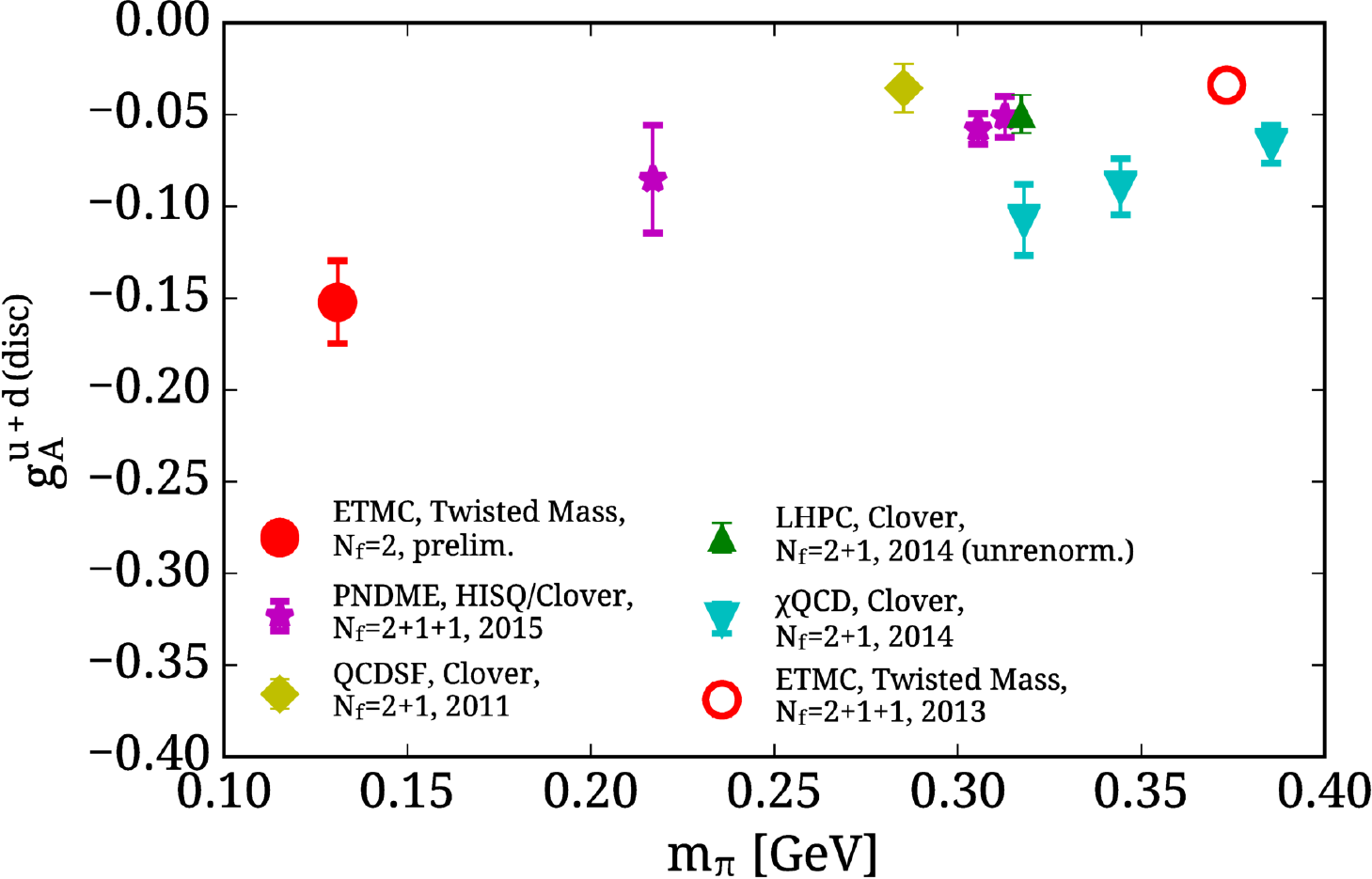}}
\vskip 0.2cm
\cl{\hspace*{0.35cm}\includegraphics[scale=0.20]{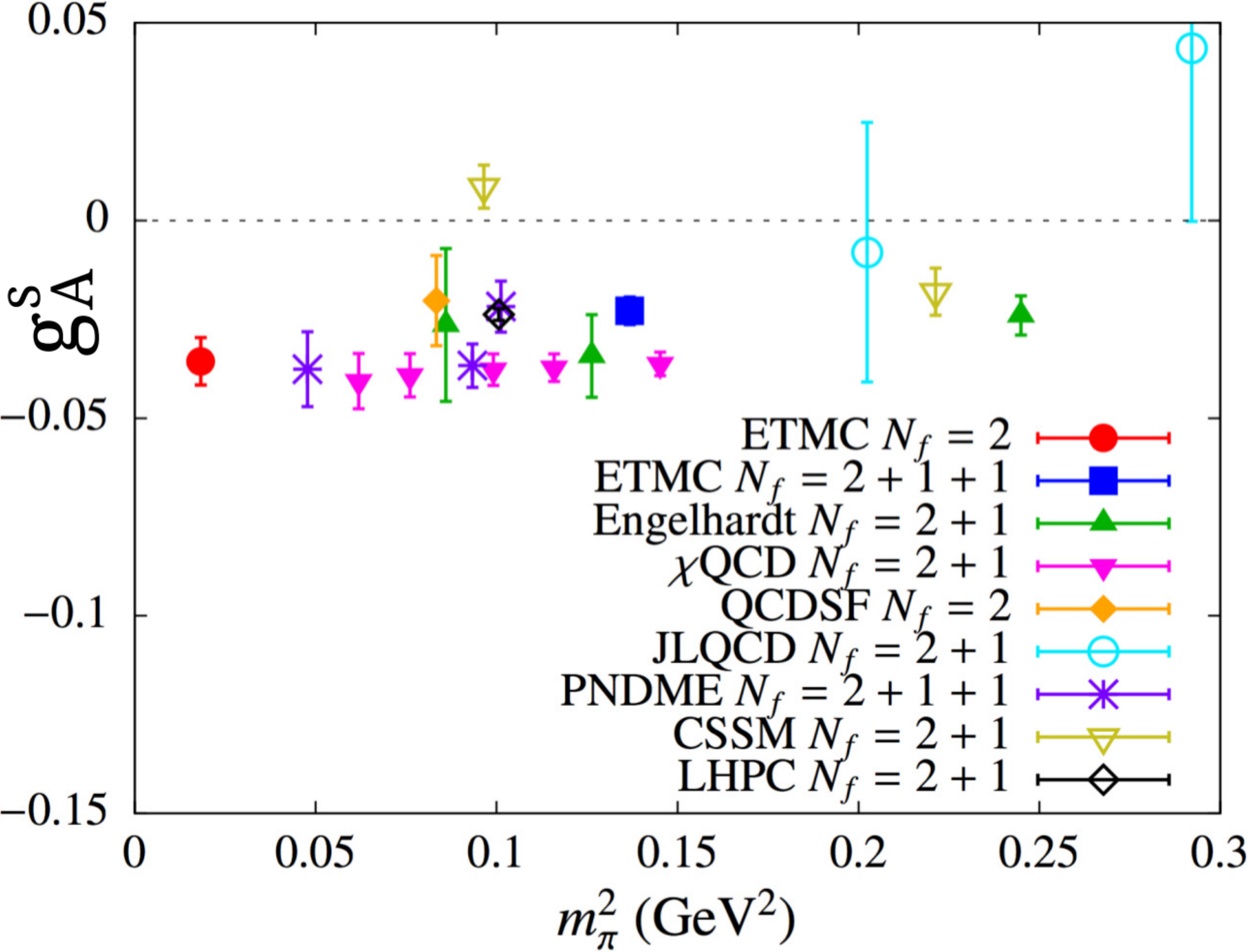}}
\vskip -0.25cm
\caption{\footnotesize{Upper plot: The disconnected light quark contribution to $g_A$ as a function of $m_\pi$. 
Lower plot: Strange contribution to $g_A$ as a function of $m_\pi^2$. The data have been taken from Refs.
\cite{Babich:2010at,QCDSF:2011aa,Engelhardt:2012gd,Abdel-Rehim:2013wlz,LHPC14,Chambers:2014tga,Yang:2015xga,Abdel-Rehim:2016pjw}.}}
\label{fig4}
\end{figure}

To give reliable estimates for the individual quark contributions to the nucleon spin, we should also take
into account the disconnected diagram (upper right of Fig.~\ref{fig1}). As already mentioned, such
contributions are more difficult to extract as they require at least an order of magnitude more statistics
compared to the connected ones, and the development of special techniques. The use of GPUs has played a
significant role in the progress of the computations for the disconnected diagram using improved actions 
with dynamical fermions. A number of results have appeared recently for the disconnected loop contributions
to $g_A$ as shown in Fig.~\ref{fig4}. We observe a nice agreement among results both for the light and 
strange quark contributions
\cite{Babich:2010at,QCDSF:2011aa,Engelhardt:2012gd,Abdel-Rehim:2013wlz,LHPC14,Chambers:2014tga,Yang:2015xga,Abdel-Rehim:2016pjw}. 
For $g_A^{light}$ we find $\sim7{-}10\%$ contribution to the spin compared to the connected part. 
Both light disconnected and strange parts are negative and thus reduce the value of $g_A^q$.
\vskip -0.25cm
\FloatBarrier

\vspace*{-.5cm}
\subsection{Axial Form Factor}
\vspace*{-.25cm}
\label{sec3sub2}

The axial form factors have attracted a lot of attention due to their relevance in experiments searching 
neutrino oscillations. Different analyses of experimental data~\cite{AguilarArevalo:2010zc,Meyer:2016oeg} 
show systematic uncertainties in the determination of the axial dipole mass, $M_A$, that are not well 
controlled due to their model dependence. Thus, lattice QCD data are vital as one can extract the $Q^2$ 
dependence of such form factors from first principle calculations.  
\vskip -0.25cm
 \begin{figure}[!h]
\cl{\includegraphics[scale=0.43]{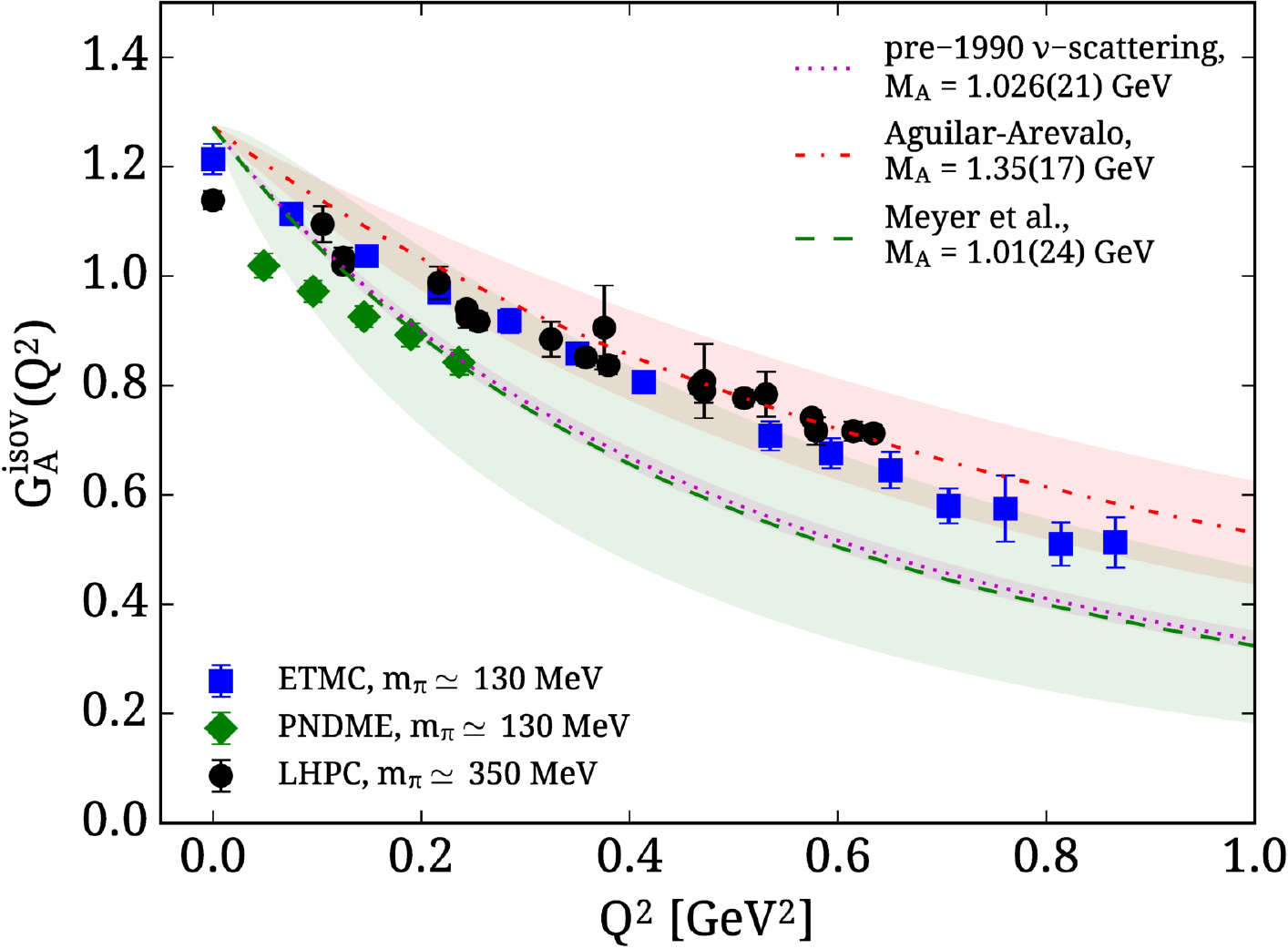}}
\vskip -0.35cm
\caption{\footnotesize{Isovector $G_A$ as a function of $Q^2$. 
The data correspond to:
$N_f{=}2$ TMF $\&$ clover by ETMC~\cite{Koutsou16},   
$N_f{=}2+{1}+{1}$ HISQ by PNDME~\cite{Jang16},   
$N_f{=}2{+}1$ clover by LHPC~\cite{Green16}. The bands show
different analysis of experimental data~\cite{AguilarArevalo:2010zc,Meyer:2016oeg}.}}
\label{fig5}
\end{figure}
\FloatBarrier
As an example we plot the axial form factor for three formulations in Fig.~\ref{fig5}
at $m_\pi{=} 130$ MeV~\cite{Koutsou16,Jang16} and 317 Mev~\cite{Green16}. Using the lattice
data one can fit to a dipole form: {\small{$G_A(Q^2) = g_A/(1+Q^2/M_A^2)^2$}}, either using the $g_A$
from the lattice data, or allowing both $g_A$ and $M_A$ to vary. In Ref.~\cite{Koutsou16} a study
of the source-sink separation shows that in such a fit, $g_A$ is approaching its experimental value
with increasing the separation, while $M_A$ is found to be consistent within error bars to the 
experimental determination of Refs.~\cite{AguilarArevalo:2010zc,Meyer:2016oeg}. 
 The value for $M_A$ extracted using TMF $\&$ clover at $m_\pi{\sim}130$MeV is $M_A{=}1.24(8)$GeV~\cite{Koutsou16} and in agreement 
with the value $M_A{=}1.24(14)$ obtained with DWF at $m_\pi{=}172$ MeV~\cite{Abramczyk:2016ziv}. PNDME reports a preliminary value of
$M_A{=}1.02(4)$GeV at $m_\pi{=}130$ MeV for the  HISQ formulation~\cite{Jang16}.
\vskip -0.3cm
 \begin{figure}[!h]
\cl{\includegraphics[scale=0.42]{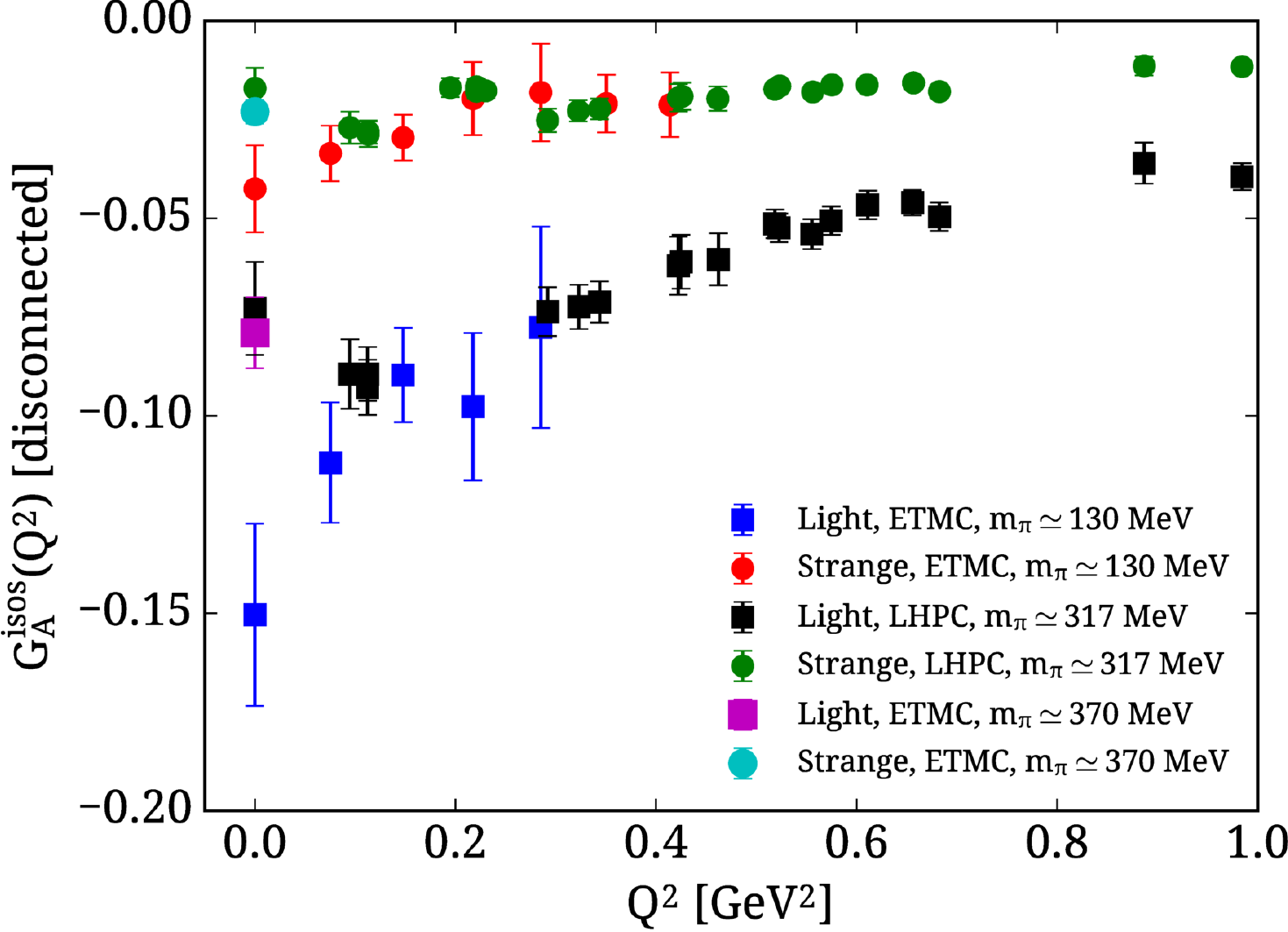}}
\vskip -0.35cm
\caption{{\footnotesize{Disconnected contribution to the axial form factor as a function of $Q^2$. 
The data correspond to $N_f{=}2$ TMF $\&$ clover by ETMC~\cite{Koutsou16}
and $N_f{=}2{+}1$ clover by LHPC~\cite{Green16}.}}}
\label{fig6}
\end{figure}
\FloatBarrier
The most recent echievement regarding the axial form factors is the computation of the 
disconnected loop contributions for the light, strange, even the charm quark; 
the latter was found to be compatible with zero~\cite{Koutsou16}. In Fig.~\ref{fig6}
we plot together the two results that appeared recently in the literature for $G_A$ obtained
with TMF $\&$ clover at the physical pion mass (ETMC~\cite{Koutsou16}) and 
clover at $m_\pi{=}317$MeV (LHPC~\cite{Green16}). For comparison, we also include
$G_A(0)$ for TMF at $m_\pi{=}375$MeV (ETMC~\cite{Abdel-Rehim:2013wlz}) which
is in agreement with the LHPC data. Despite the difference in the pion mass of the 
ensembles, we find a good agreement for $Q^2{>}0.1$GeV$^2$. 
However, for small values of $Q^2$ the data at the physical point exhibit a steep 
downward trend. 

\vspace*{-0.4cm}
\subsection{Quark Momentum Fraction}
\vspace*{-0.1cm}
\label{sec3sub2}

\begin{figure}[!h]
\cl{\includegraphics[scale=0.23]{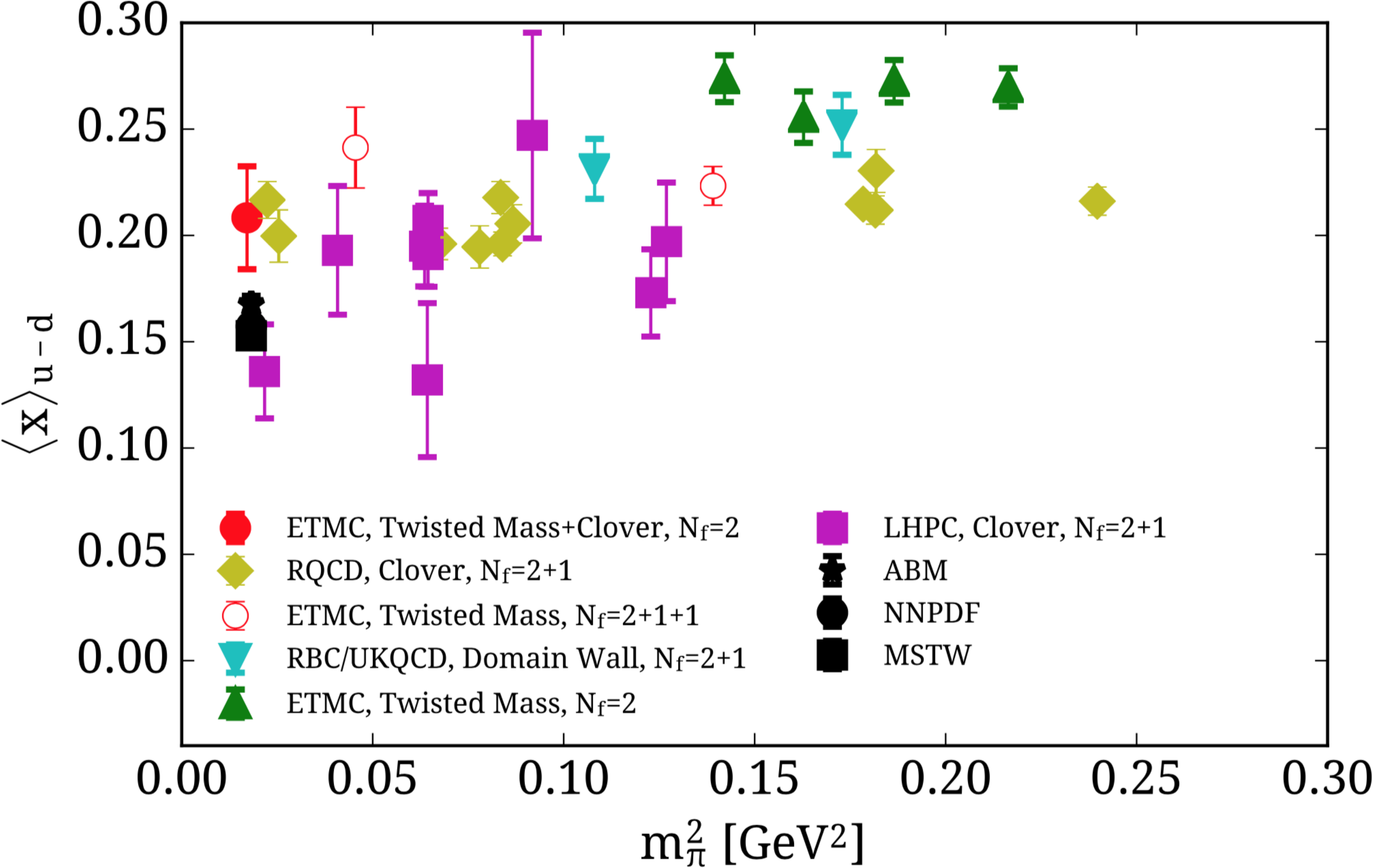}}
\vskip -0.3cm
\caption{\footnotesize{Recent results on $\langle x \rangle_{u{-}d}$, as a function 
of $m_\pi^2$. Data correspond to: 
$N_f{=}2$ TMF (ETMC~\cite{Alexandrou:2011nr}), 
$N_f{=}2{+}1$ DWF (RBC/UKQCD~\cite{Aoki:2010xg}),
$N_f{=}2$ Clover (RQCD~\cite{Bali:2014gha}),
$N_f{=}1{+}2$ Clover (LHPC~\cite{Green:2012ud}), 
$N_f{=}2{+}1{+}1$ TMF (ETMC~\cite{Alexandrou:2013joa}),
$N_f{=}2$ TMF with Clover (ETMC~\cite{Alexandrou:2016tuo}). }}
\label{fig7}
\end{figure}

Another observable that contributes to the nucleon spin is the quark momentum
fraction, $\langle x\rangle_q {\equiv} A^q_{20}(0)$,  as shown in Eqs.~(\ref{eq1}) - (\ref{eq2}). 
$\langle x\rangle_q$  is extracted from the one-derivative vector
current, and it is a scheme and scale dependent quantity.
Fig.~\ref{fig7} shows results on $\langle x \rangle_{u{-}d}$ converted to $\overline{\rm MS}$ scheme at
a scale of 2~GeV, and overall, the lattice data overestimate the phenomenological values. The phenomenological estimates 
extracted from different analyses (Refs. [56-61] of Ref.~\cite{Alexandrou:2013joa}) show deviation, which, 
however, is significantly smaller than the discrepancy shown with the lattice data. The lattice results of ETMC 
and RQCD close to the physical point are in agreement, while the LHPC point at ${\sim}150$MeV is close to 
the experimental point. It is interesting to note that removal of excited states has been applied to the LHPC point,
thus, comparison with the other lattice data is not meaningful.
\vskip -0.3cm
\FloatBarrier
\vskip -0.6cm
\begin{figure}[!h]
\cl{\includegraphics[scale=0.45]{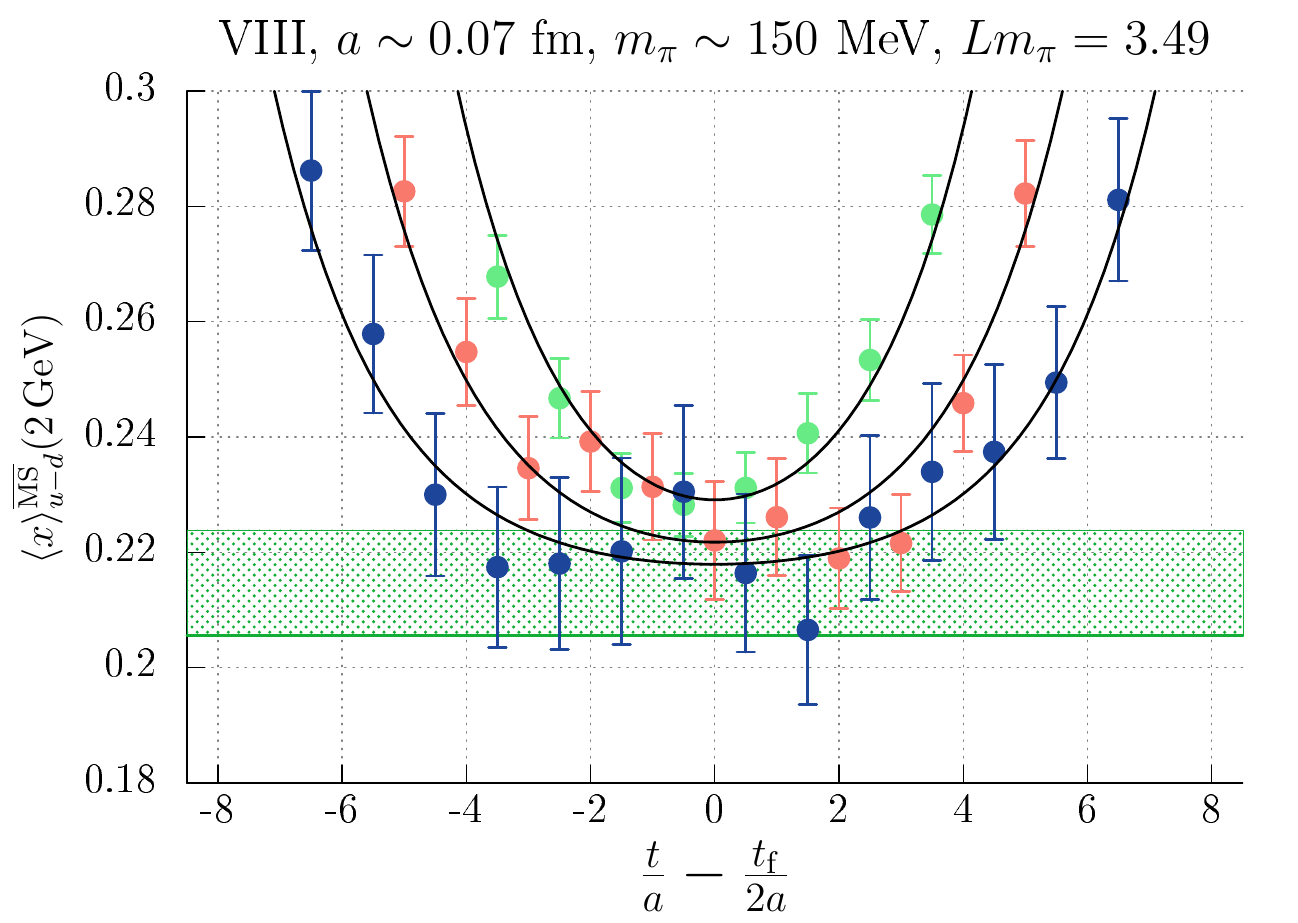}}
\vskip -0.3cm
\caption{\footnotesize{The ratio leading to $\langle x \rangle_{u-d}$, as a function of the current
time insertion, for $T_{sink}{=}8a,\,11a,\,14a$. The green dashed band indicates
a two-state combined fit at various $T_{sink}$.}}
\label{fig8}
\end{figure}
\FloatBarrier
\vspace*{-0.3cm}
The data of Fig.~\ref{fig7} correspond to source-sink separation of $T_{sink}{\sim}$ 1-1.2 fm, 
which might not be large enough. A number of studies were undertaken to understand excited
state effects and data from different formulations show a large contamination. In particular, 
by increasing the source-sink separation one observes a decrease of the extracted $\langle x \rangle_{u-d}$
about ${\sim}10\%$. As an example we show the investigation by RQCD~\cite{Bali:2014gha} at $m_\pi=150$MeV,
and in Fig.~\ref{fig8} we demonstrate the ratio leading to $\langle x \rangle_{u-d}$ at three separations, with
maximum value ${\sim}1$fm. It should be noted that for this observable a separation of at least 1.5fm is necessary
to extract reliable results~\cite{Dinter:2011sg,Abdel-Rehim:2015owa}.

\begin{figure}[!h]
\cl{\includegraphics[scale=0.33]{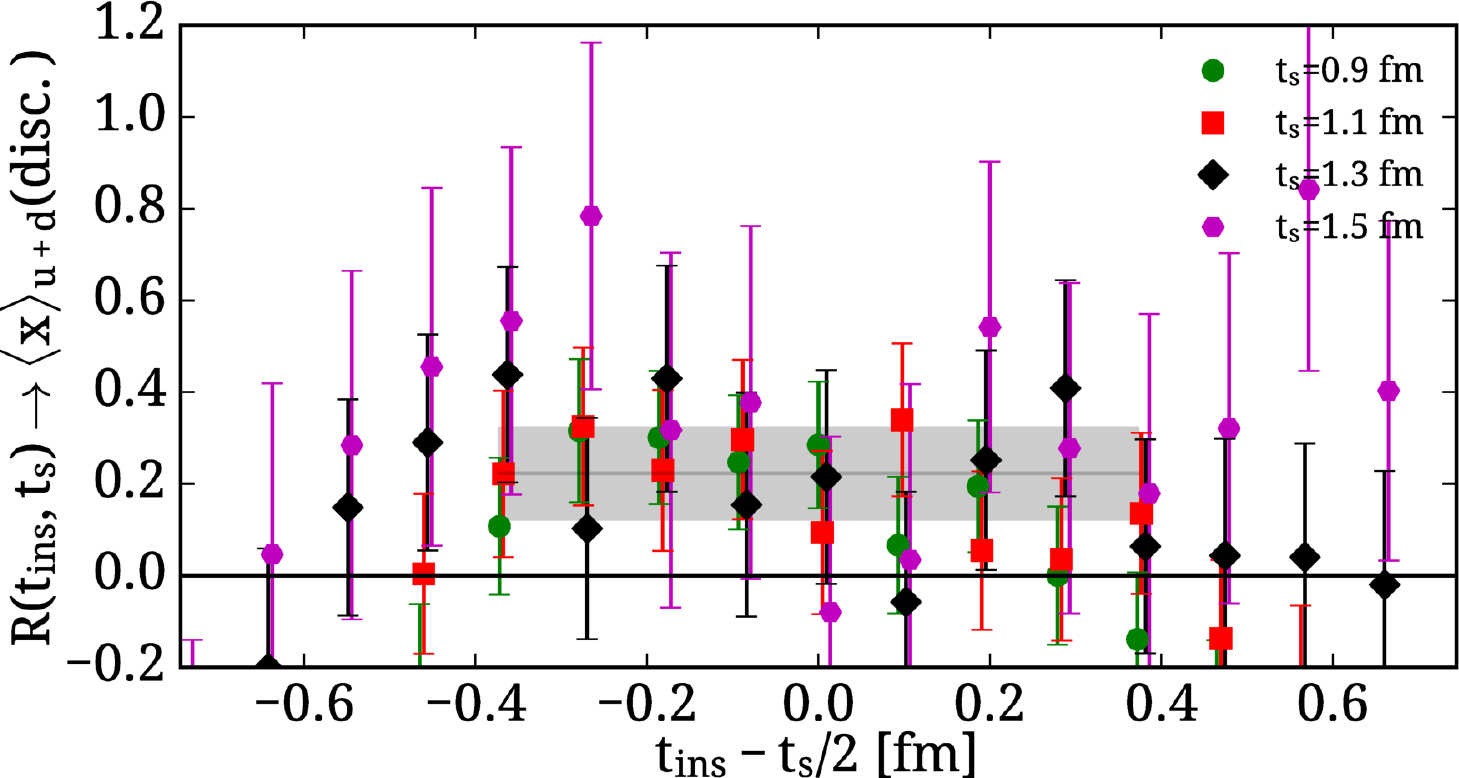}
\includegraphics[scale=0.185]{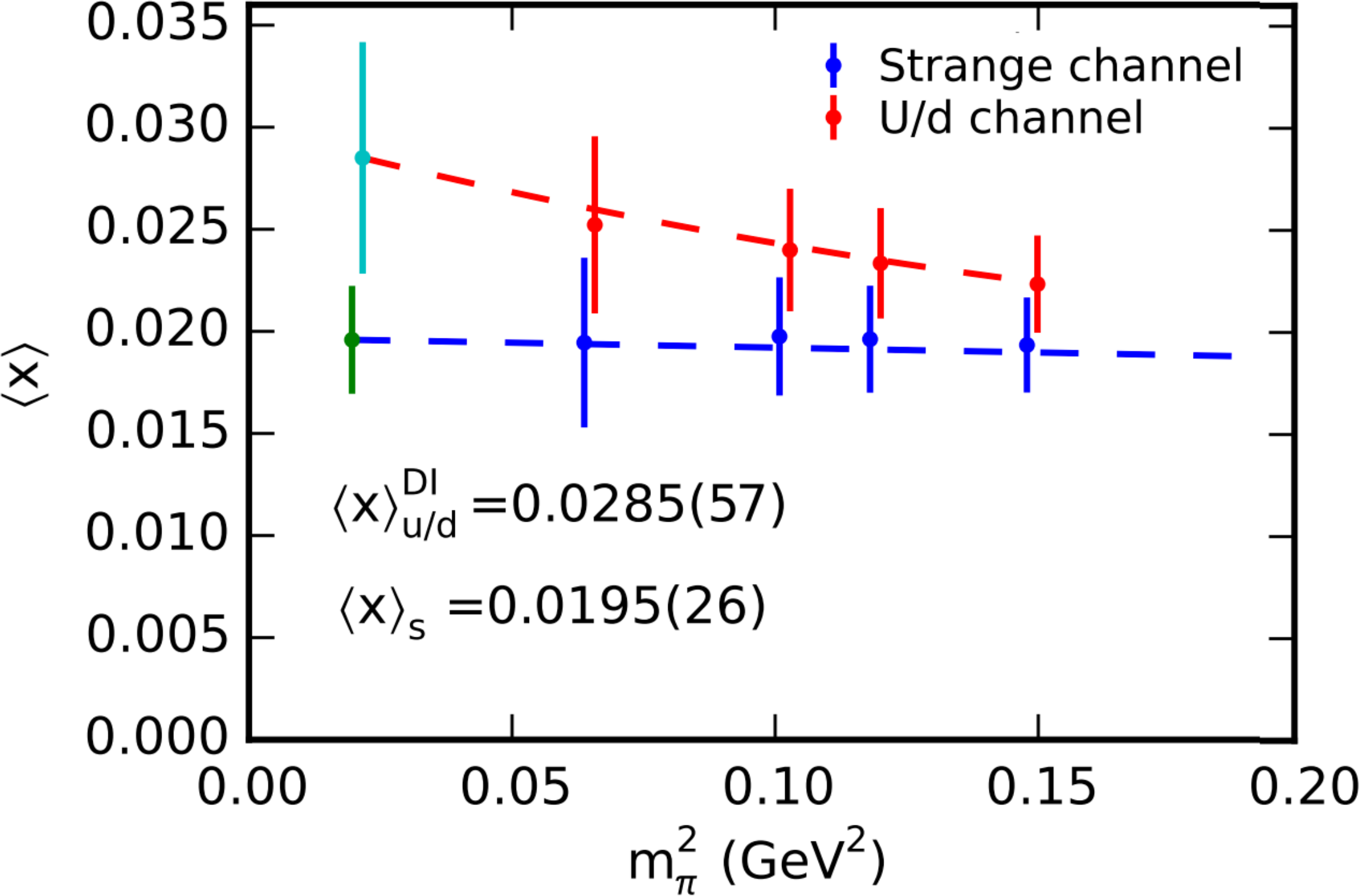}}
\vskip -0.25cm
\caption{\footnotesize{Left: Ratio for the disconnected $\langle x \rangle_{u{+}d}$ as a function of the current 
insertion time and for $T_{sink}{=}$0.9{-}1.5 fm using TMF fermions at $m_\pi{=}130$MeV~\cite{Abdel-Rehim:2016pjw}. 
Right: Light disconnected and strange $\langle x \rangle$ for several ensembles from $\chi$QCD~\cite{Sun:2015pea}.}}
\label{fig9}
\end{figure}

An important component in the nucleon spin is the disconnected contributions to $\langle x \rangle$, and 
recently there has been a computation directly at the physical point~\cite{Abdel-Rehim:2016pjw}, which shows
a significant contribution from the light quarks:  $\langle x \rangle^{DI}_{u{+}d}{=}0.223(99)$, and a non negligible
for the strange: $\langle x \rangle^{DI}_s{=} 0.092(41)$. At heavier pion masses, the disconnected contributions 
for this quantity were found to be small~\cite{Abdel-Rehim:2013wlz}, but it is expected to have a stronger pion mass
dependence as one approaches the physical point. This is also demonstrated in the right panel of Fig.~\ref{fig9} where
there is an upward trend with decreasing $m_\pi$~\cite{Sun:2015pea}.
\vskip -0.2cm
\FloatBarrier
It is worth mentioning that the ratio $\langle x \rangle_s/\langle x \rangle_{u/d}$ is consistent
between lattice data and the global analysis of Ref.~\cite{Martin:2009iq}, provided that the `disconnected sea' 
contribution dominates small $x$ and that $(s {+} \bar s)/(\bar u {+} \bar d)$ is relatively flat in the
small $x$ region. In particular, $\langle x \rangle_s/\langle x \rangle_{u/d} {=} 0.76(30)$ (ETMC at $m_\pi{=}130$MeV), 
$\langle x \rangle_s/\langle x \rangle_{u/d} {=} 0.78(03)$ ($\chi$QCD chirally extrapolated).

\vspace*{-.25cm}
\section{Gluon Momentum Fraction}
\vspace*{-.3cm}
\label{sec4}

To complete the picture of the nucleon spin one must consider contributions from the gluons. 
Gluon contributions are poorly known from lattice QCD, as they require a disconnected insertion, 
have low signal quality and exhibit operator mixing~\cite{Caracciolo:1991cp}.  
Until recently, the only results for the gluon momentum fraction, $\langle x \rangle_g$ were 
quenched~\cite{Horsley:2012pz,Liu:2012nz}, and an alternative investigations using the 
Energy-Momentum Tensor decomposition is presented in Ref.~\cite{Yang:2016plb}. 
In these proceedings we highlight the only computation at the physical point in which the mixing between
$\langle x \rangle_g$ and $\langle x \rangle_q$ has been eliminated~\cite{Alexandrou:2016ekb}; 
this is a test for the momentum sum rule, $\sum_q \langle x \rangle_q {+} \langle x \rangle_g {=} 1$, 
and may shed light on the proton spin puzzle.
\vskip -0.3cm
\begin{figure}[!h]
\cl{\includegraphics[scale=0.11]{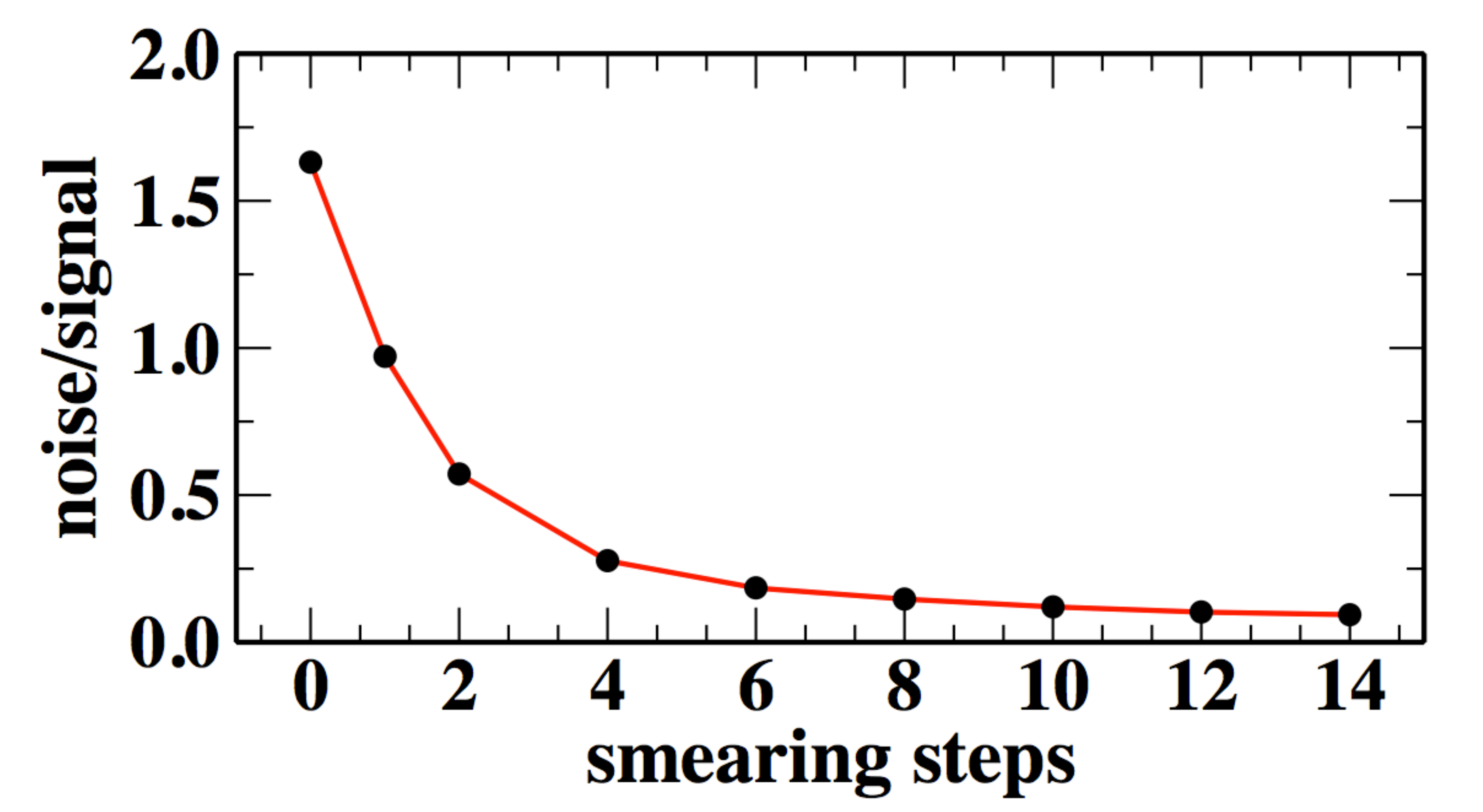}
\includegraphics[scale=0.19]{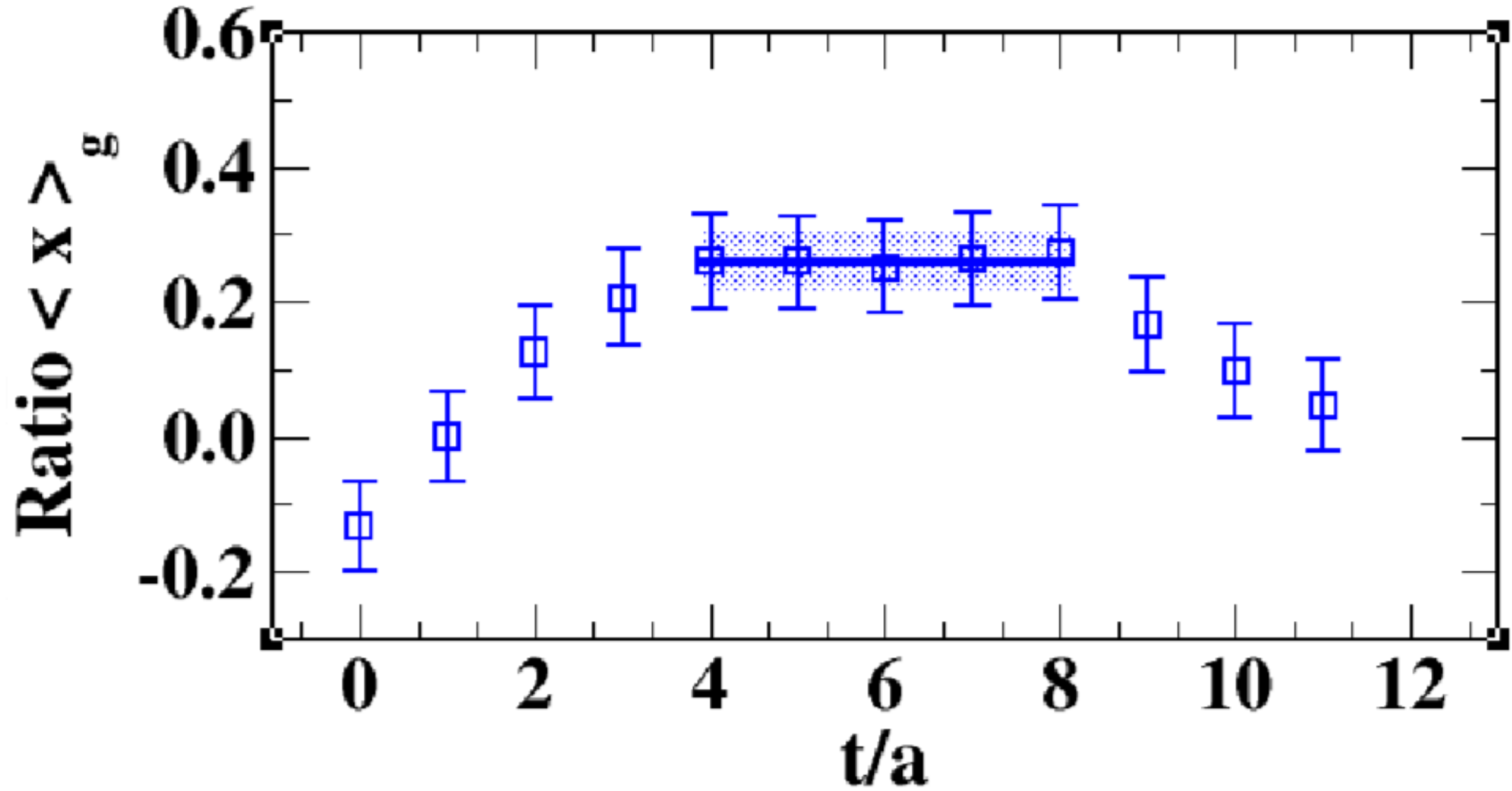}}
\vskip -0.25cm
\caption{\footnotesize{Left: noise-to-signal ratio versus the smearing steps.
Right: Ratio for $\langle x \rangle_g$ for TMF at $m_\pi{=}130$MeV~\cite{Alexandrou:2016ekb}.}}
\label{fig10}
\end{figure}
\FloatBarrier
Although disconnected contributions are notoriously difficult and noisy, applying smearing to the gauge links
in the gluon operator improves the quality of the signal. This was demonstrated in Ref.~\cite{Alexandrou:2016ekb} 
where increasing the smearing steps leads to a significant reduction of the noise-to-signal ratio, as seen in the left
plot of Fig.~\ref{fig10}. As a result a good quality of plateau for $\langle x \rangle_g$ is achieved (right plot of
Fig.~\ref{fig10}).

Following the renormalization prescription outlined in Ref.~\cite{Alexandrou:2016ekb} and the total light and strange quark
momentum fraction, it is found that the final renormalized value for the gluon momentum fraction is $\langle x \rangle_g{=}0.273(23)(24)$.
Using all quark and gluon contributions to the nucleon momentum the sum rule is then satisfied: 
$\sum_q \langle x \rangle_q {+} \langle x \rangle_g {=} 1.01(10)(2)$.

\vskip -0.5cm
\section{Proton Spin}
\label{sec5}
\vskip -0.3cm

\begin{figure}[!h]
\cl{\includegraphics[scale=0.165]{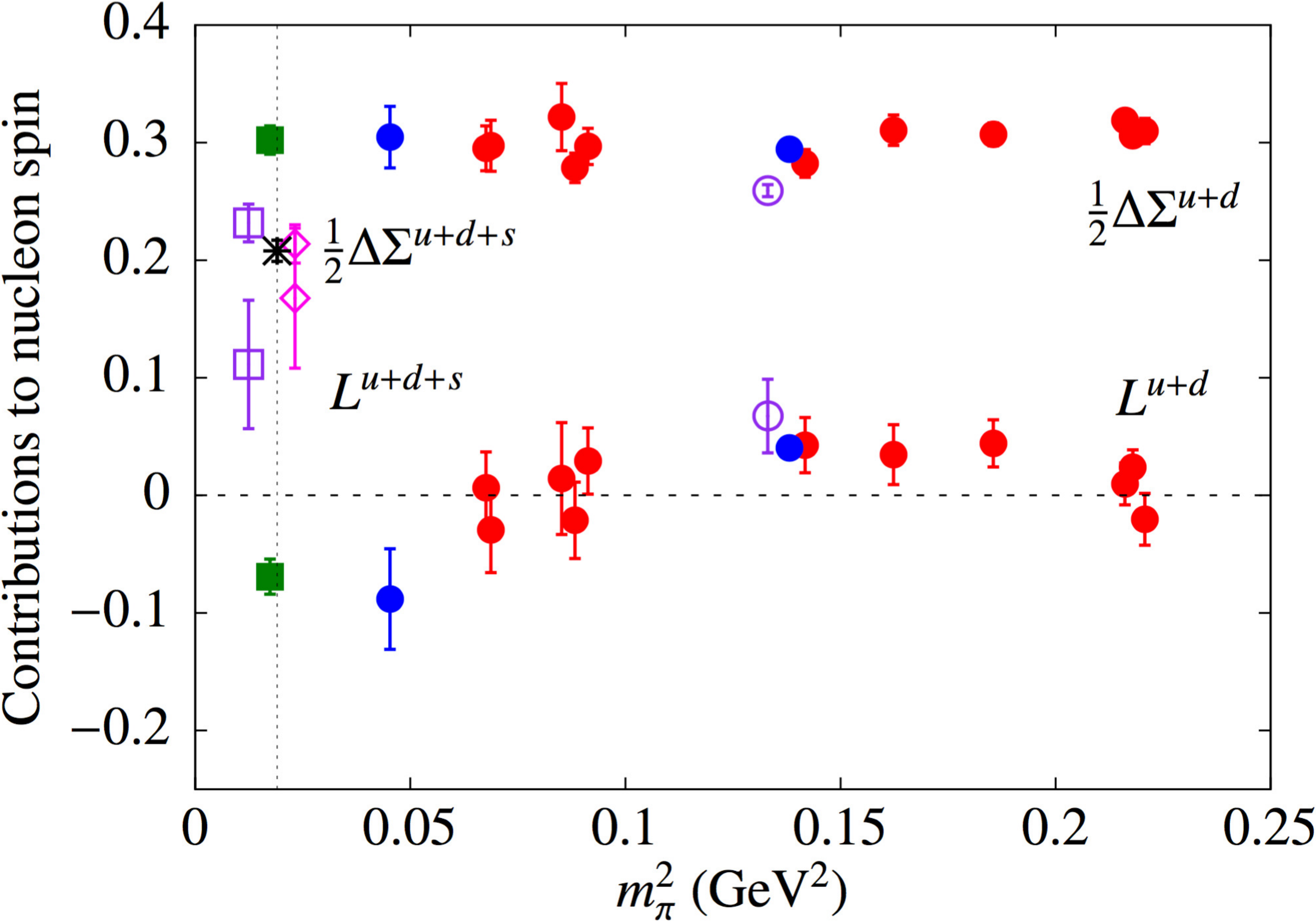}}
\vskip -0.25cm
\caption{\footnotesize{The intrinsic spin $\Delta\Sigma_{u{+}d{+}s}$ and orbital angular 
momentum $L^{u{+}d{+}s}$} for TMF.}
\label{fig11}
\end{figure}

Using the decomposition presented in Eqs.~(\ref{eq1}) - (\ref{eq2}) one can obtain from lattice data
the total quark spin $J^q$ and the intrinsic spin $\Delta\Sigma$. Furthermore, using the spin sum
rule $J^q{=}\Delta\Sigma^q/2 {+} L^q$ one can extract lattice estimates on the orbital angular 
momentum $L^q$. In the left panel of Fig.~\ref{fig11} we plot the total $\Delta\Sigma^q$ (upper points), 
and $L^q$ (lower points) using TMF~\cite{Alexandrou:2011nr,Alexandrou:2013joa,Alexandrou:2016ekb}.
The open symbols correspond to two ensembles at $m_\pi{=}130,\,375$MeV, for which the disconnected light 
and strange quarks have been included. Focussing on the data at the physical pion mass, where the green 
filled squares are the connected $u{+}d$, one observes that the addition of the disconnected $u,\,d,\,s$ shifts 
$\Delta\Sigma$ towards the experimental point (black star). The connected contribution to
$L^{u{+}d}$ is very small and the inclusion of the disconnected contributions gives a positive value of $\sim 0.15$.
\vskip -0.2cm
\FloatBarrier
\vskip -0.05cm
\textit{Acknowledgments:}
 {\small{I would like to thank all members of the ETMC for a fruitful collaboration and in particular 
 C. Alexandrou, A. Abdel-Rehim, K. Jansen, K. Hadjiyiannakou, Ch. Kallidonis,
G. Koutsou, H. Panagopoulos and A. Vaquero for their invaluable contributions.}}

\bibliography{references}
\end{document}